\documentclass[fleqn,usenatbib]{mnras}

\usepackage[T1]{fontenc}
\usepackage{ae,aecompl}

\usepackage{graphicx}	
\usepackage{amsmath}	
\usepackage{amssymb}	
\usepackage{nccmath}
\usepackage{subfig}
\usepackage{soul}
\usepackage{enumitem}

\title[Can jets make radioactive emission bluer?]{Can jets make the radioactively powered emission from neutron star mergers bluer?}

\author[L. Nativi et al.]{
L. Nativi$^{1}$\thanks{E-mail: lorenzo.nativi@astro.su.se},
M. Bulla$^{2}$, 
S. Rosswog$^{1}$, C. Lundman$^{1}$, G. Kowal$^{3}$, \and
D. Gizzi$^{1}$, G.~P. Lamb$^4$ and A. Perego$^{5,6}$
\\
$^{1}$Department of Astronomy and Oskar Klein Centre, Stockholm University, AlbaNova 10691 Stockholm, Sweden\\
$^{2}$Nordita, KTH Royal Institute of Technology and Stockholm University, Roslagstullsbacken 23, SE-106 91 Stockholm, Sweden\\
$^{3}$Escola de Artes, Ci\^{e}ncias e Humanidades, Universidade de S\~{a}o Paulo, Av. Arlindo B\'{e}ttio, 1000-Vila Guaraciaba,\\ CEP: 03828-000, S\~{a}o Paulo-SP, Brazil\\
$^{4}$School of Physics and Astronomy, Leicester University, University Road, Leicester LE1 7RH, UK\\
$^{5}$Dipartimento di Fisica, Universit\`{a} di Trento, Via Sommarive 14, 38123 Trento, Italy\\
$^{6}$Istituto Nazionale di Fisica Nucleare, Sezione di Milano-Bicocca, Piazza della Scienza 20100, Milano, Italy
}

\date{Accepted XXX. Received YYY; in original form ZZZ}

\pubyear{2020}

\begin{document}
\label{firstpage}
\pagerange{\pageref{firstpage}--\pageref{lastpage}}
\maketitle

\begin{abstract}

Neutron star mergers eject neutron-rich matter in which heavy elements are synthesised. The decay
of these freshly synthesised elements powers electromagnetic transients 
(``macronovae"
or ``kilonovae") whose luminosity and colour strongly depend on their nuclear composition.  If the
ejecta are very neutron-rich (electron fraction $Y_\mathrm{e} < 0.25$), they contain fair amounts of lanthanides and actinides which
have large opacities and therefore efficiently trap the radiation inside the ejecta so that the  emission
peaks in the red part of the spectrum. Even small amounts of this high-opacity material can obscure
emission from lower lying material and therefore act as a ``lanthanide curtain".  Here, we investigate
how a relativistic jet that punches through the ejecta can potentially push away a significant fraction
of the high opacity material before the macronova begins to shine. We 
use the results of detailed neutrino-driven wind studies as initial conditions and explore 
with 3D special relativistic hydrodynamic simulations how jets are
propagating through these winds.
Subsequently, we perform Monte Carlo radiative transfer calculations to explore the  resulting
macronova emission. We find that the hole punched by the jet makes the macronova brighter and
bluer for on-axis observers during the first few days of emission, and that more powerful jets have
larger impacts on the macronova.
\end{abstract}




\vspace*{5cm}

\section{Introduction}
\label{sec:intro}
 The first joint detection of gravitational 
 and electromagnetic waves from a binary neutron star merger 
 on 17 August 2017
 marked the beginning of a new era of astrophysics  
 \citep{Abbott2017a, Abbott2017b,Abbott2017c}. About two seconds after
 the peak of the gravitational wave (GW) signal, a short gamma-ray burst (sGRB) was detected \citep{Goldstein2017,savchenko17} and the remarkable event was followed up during the subsequent days and weeks all across the electromagnetic spectrum, starting with early UV, optical and IR signals \citep[e.g.][]{Arcavi2017,Cowperthwaite2017,Evans2017,Drout2017,Kasliwal2017Sc,Pian2017,Smartt2017,SoaresSantos2017,Utsumi2017} later
 followed by X-rays \citep{Troja2017, Davanzo2018, Margutti2018} and radio emission \citep{Alexander2017, Hallinan2017, Troja2019}. 
 The UVOIR signal, observed from the first day up to two weeks, was broadly consistent with the emission expected from a macronova (or kilonova, hereafter MKN) \citep{Kasen2017,Perego2017,Tanaka2017, Rosswog2018}, 
 a thermal transient powered by the radioactive decay of the freshly synthesized r-process elements \citep{li98, kulkarni05, rosswog05, metzger10,metzger17}.
The emission in X-rays and the radio band was interpreted as 
being produced by a relativistic jet, observed slightly off-axis, 
interacting with previously ejected matter 
\citep{Alexander2017, Margutti2017, Hallinan2017, Kasliwal2017Sc, Lazzati2018, Margutti2018, Mooley2018, Lamb2018, Kathir2018, Ghirlanda2019, Davanzo2018}. The origin of the sGRB is so far unknown, but could plausibly have been produced inside the jet, or in a shock breakout event when the jet emerges from the ejecta \citep{Nakar2017, Lamb2017,Gottlieb2018a, Gottlieb2018b, Beloborodov2018}.

Soon after the discovery of the first neutron 
star binary \citep{hulse75} it was realized
that such binaries would be driven through GW
emission towards  a violent collision which could potentially
eject neutron-rich matter \citep{lattimer74}. It remained, however, an open question for more than two decades
whether such extremely gravitationally bound 
systems can actually eject any mass at all.
The first hydrodynamic-plus-nucleosynthesis calculations
\citep{rosswog98b,rosswog99,freiburghaus99b} showed  
that $\sim 0.01$ M$_\odot$ of neutron-rich matter
is dynamically ejected during a merger and the  nuclear network calculations demonstrated that 
the extremely neutron-rich ejecta effortlessly reproduce
the elements up to and beyond the 3rd r-process/``platinum" peak (A=195) \citep{freiburghaus99b}. This
had been a major challenge for all other suggested r-process production
sites. These results immediately triggered the discussion on 
how a neutron star merger would appear electromagnetically \citep{li98}.

While these early studies demonstrated the viability of
neutron star mergers as a major r-process site, they identified
only one ejection channel: ``dynamical ejecta" which  are tidally flung 
out by gravitational torques. Since they are never  substantially heated, these ejecta carry their original
$\beta-$equilibrium electron fraction from the 
original neutron star, $Y_{\rm e} \approx 0.05$,  and this enormous
neutron-richness allows them to undergo a ``fission cycling"
process \citep{goriely11a,korobkin12a} which produces a very
robust r-process abundance distribution close to the solar pattern
for $A \geq 130$, but hardly any lighter r-process
elements. \cite{oechslin07a} pointed out that there is a second channel of mass ejection
that also happens on a dynamical time scale: shock-heated matter from the
interface where the stars come into contact. As of today, 
many more mass ejection channels have been discussed: matter 
that becomes unbound on secular time scales
($\sim 1$ s) from the post-merger accretion torus 
\citep{metzger08a,beloborodov08,fernandez13b,fernandez15,just15,siegel17a,siegel18,miller19,fernandez19}, 
as MHD-driven winds  \citep{siegel15} and by viscous effects \citep{shibata17,radice18a,shibata19a} from a long-lived neutron 
star merger remnant. Similar to the case of proto-neutron stars, 
the enormous neutrino luminosities ($> 10^{53}$ erg s$^{-1}$) after 
a neutron star merger can also drive substantial matter outflows
\citep{ruffert97a,rosswog02b,dessart09,Perego2014,Martin2015,radice18b}. 
The secular torus ejecta contain approximately 40\% of the initial
torus mass and, although the latter may vary substantially from
case to case, they likely contribute the lion's share to the total
ejecta mass. Due to their different thermal histories and
exposure times to neutrinos, the ejecta channels can have
different electron fractions $Y_{\rm e}$ and therefore 
different nucleosynthesis yields\footnote{For the low-entropy ejecta
($s \lesssim 50 k_{\rm B}/{\rm baryon}$) of a neutron star merger
the electron fraction is the most crucial parameter for the nucleosynthesis.}. For electron fractions
below a critical value $Y_{\rm e}^{\rm crit}\approx 0.25$ \citep{korobkin12a,lippuner15} lanthanides 
and actinides are efficiently produced which, due their open f-shells, have particularly high bound-bound 
opacities \citep{kasen13a,barnes13,tanaka13,tanaka20} and therefore lead to red transients that peak days after the merger. Ejecta with electron fractions above $Y_{\rm e}^{\rm crit}$, 
in contrast, only produce ``lighter'' elements with lower opacities
and thus result in bluer transients that peak after about one day.
Opaque, low-$Y_{\rm e}$ ejecta blocking the view on high-$Y_{\rm e}$ ejecta
can lead to a ``lanthanide curtaining" effect 
\citep{kasen15, Wollaeger2018} which will efficiently block blue
light. Therefore it is important to understand the layering, 
dynamics, interaction and potential mixing of different ejecta channels.

The multi-messenger detection of GW170817 provided evidence that neutron star mergers can produce short GRBs\footnote{However, GRB170817A was extremely under-luminous, and the favoured emission models do not consider the gamma-rays to originate directly from the jet, implying that GRB170817A was no ordinary short GRB.}. Given the expected complexity of the matter distribution engulfing the remnant,
it is interesting to understand under which conditions a relativistic
jet can  successfully drill through the ejecta cloud \citep{Murguia-berthier2014,Beniamini2020}
and whether/how it affects the mixing/interaction between the different
components. This could have substantial consequences
for the layering and interaction of the ejecta and it can have potentially large effects on the ``lanthanide
curtaining".

Here we explore the hydrodynamic 
interaction of a relativistic jet
with previously launched neutrino-driven winds from a long-lived 
neutron star merger remnant. Contrary to earlier studies \citep{Zhang2003, Mizuta2009, Mizuta2013, Nagakura2014,  Murguia-berthier2014, Duffell2018, Harrison2018}
we use actual simulation results \citep{Perego2014} as initial conditions for the 
surrounding wind structure and dynamics. We are 
particularly interested in the question how jets of 
different power impact the
observable MKN broad-band light curves. The light curves are 
obtained by running 3D Monte-Carlo radiative transfer simulations
on a homologously expanding matter background. For this matter
background we use the results of our wind-plus-jet simulations
and we add an additional component that is meant to represent the
likely  important secular disk ejecta which cannot be 
modelled self-consistently together with our large-scale jet simulation. 
The jet interaction with the ejecta produces an additional contribution, the cocoon emission. While propagating the jet inflates a pressurised cocoon that leads to an additional electromagnetic signal on similar time scales as MKN, and over a relatively wide range of viewing angles \citep{Gottlieb2018a}. This contribution is  not considered in the present work.
\par Our paper is 
structured as follows. We begin in Section~\ref{sec:methodology} 
with an overview of our numerical methods and briefly describe the relativistic adaptive mesh-refinement hydrodynamics
code \textsc{amun}  and the radiative transfer code \textsc{possis}. Our simulation setup
is explained in Section~\ref{sec:sim_setup} and we present and discuss 
our results in Section~\ref{sec:results}. A concise summary is offered
in Section~\ref{sec:summary}.

\section{Methodology}
\label{sec:methodology}

\subsection{Special-relativistic AMR hydrodynamics}
\label{sec:srhd}
The evolution of an ideal, relativistic fluid is governed 
by the conservation of baryon number and four-momentum. The
corresponding equations form a set of hyperbolic equations
which can be written as:
\begin{ceqn}
\begin{equation}
\label{eq:conservedeq}
 \frac{\partial\mathbf{u}}{\partial t} +
 \frac{\partial\mathbf{F}^i(\mathbf{u})}{\partial x^i} = 0. 
\end{equation}
\end{ceqn}
These equations are solved for a set of six conserved variables $\mathbf{u} = (D, S^i, E, X_{\rm a})^T$, which are related to the physical variables proper rest-mass density, velocity, pressure and passive scalar $\mathbf{q}=(\rho, v^i, p, x_{\rm a})^T$ by the relations
\begin{equation}
    D          = \Gamma\rho\ , \ 
    \mathbf{S} = \Gamma^2\rho h \mathbf{v}\ , \
    E          = \Gamma^2\rho h - p - D\ , \
    X_{\rm a}  = D x_{\rm a},
\end{equation}
where $\Gamma=(1-v^iv_i)^{-1/2}$ is the local Lorentz factor, $h$ the specific enthalpy and we have used units in which the speed of light $c=1$.
The system is closed by assuming an adiabatic equation of state (EoS) $p\propto\rho^{\gamma}$ for which
we use an ideal gas law so that the specific enthalpy reads
\begin{ceqn}
\begin{equation}
\label{eq:specific_enth}
h(\rho,p) = 1+\frac{\gamma}{\gamma-1}\frac{p}{\rho}.
\end{equation}
\end{ceqn} 
In the current set of simulations the specific heat ratio is assumed to take the constant value for a radiation dominated gas $\gamma=4/3$.
\par We perform this study with \textsc{amun}
(\url{https://gitlab.com/gkowal/amun-code}), a parallel, special-relativistic,
Eulerian (magneto-)hydrodynamics   code. The
evolution scheme follows a Godunov approach based on an oct-tree
hierarchical Cartesian structured grid with adaptive mesh refinement
\citep{Quirk1991, DeZeeuw1993}. We further
use a 3rd order Strong Stability Preserving Runge-Kutta 
time
integration algorithm \citep{Gottlieb2011} with the 
Courant-Friedrichs-Lewy (CFL) parameter \cite{press92} set to 0.5, a second order
TVD reconstruction with the MinMod
limiter \citep{Toro1999}, and an HLL Riemann
solver \citep{Harten1983b} to compute the fluxes between adjacent
cells.

\subsection{Radiative transfer with the POSSIS code}
\label{sec:possis}
Broad-band light curves for the models investigated in this study are calculated with the 3-D time-dependent Monte Carlo radiative transfer code \textsc{possis} \citep{Bulla2019b}. Assuming homologous expansion for the ejected material, \textsc{possis} simulates Monte Carlo photons propagating throughout the expanding ejecta and interacting with matter via either electron scattering or bound-bound interactions (bound-free and free-free processes are subdominant at the wavelengths investigated in this study, \citealt{Tanaka2018}). Synthetic observables including spectral energy distributions (SEDs) and light curves can be extracted for different viewing angles defined by their polar ($\theta_{\rm obs}$) and azimuthal ($\phi_{\rm obs}$) angles (where $z$ is the jet direction and $xy$ is the orbital plane). We use analytic functions based on state-of-the-art calculations \citep{Tanaka2018} for the wavelength- and time-dependence of opacities \citep{Bulla2019b}. Compared to simulations in \citet{Bulla2019b}, we adopt an improved version of \textsc{possis} where the temperature is no longer parameterized, but rather estimated from the mean intensity of the radiation field at each time and in each zone. We refer the reader to \citet{Bulla2019b} for more details about the \textsc{possis} code, more details about our simulations follow in Section~\ref{sec:radsetup}.

\section{Simulation setup}
\label{sec:sim_setup}
\subsection{Relativistic hydrodynamics}
\begin{figure*}
    \centering
	\includegraphics[width=0.8\textwidth]{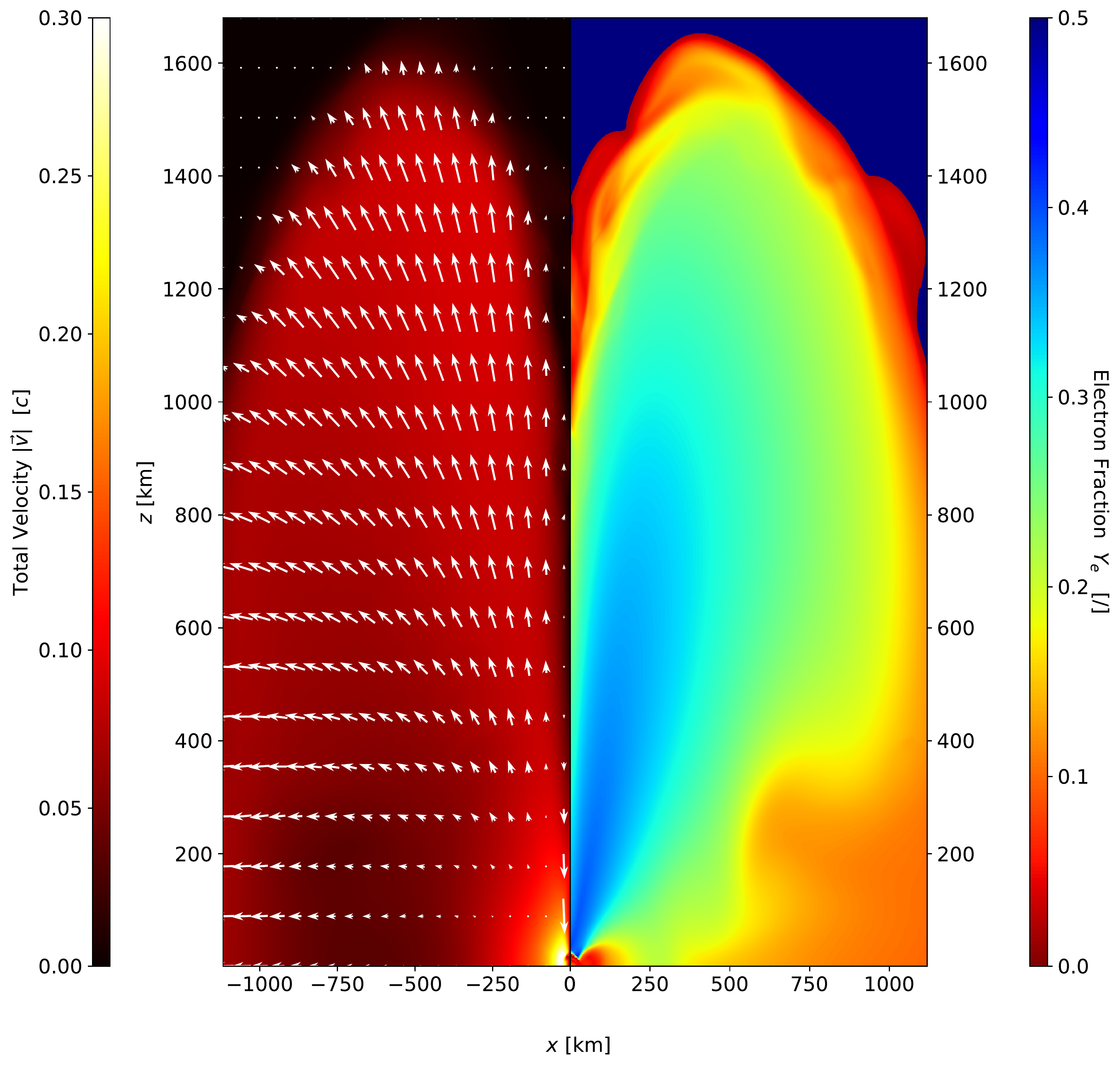}
    \caption{Vertical slices that show our wind initial conditions based on \citet{Perego2014}. The velocity is
    shown in the left and the electron fraction distribution in the right panel. 
    The total velocity is dominated by the rotational component in the inner disk, 
    while the white vectors show the azimuthal components of the velocity field. While 
    most of the wind is regularly expanding in a wing-shaped configuration, a downflow is present along the rotation axis. The electron fraction shows a fast expanding lanthanide-poor region ($Y_{\rm e}>0.25$) surrounded by a thin layer
    of low-$Y_{\rm e}$ material ($Y_{\rm e}<0.25$).}
    \label{fig:windmaps}
\end{figure*}

As initial conditions we use a simulated
neutrino-driven wind model from
\cite{Perego2014}. At $\approx 105\ \rm{ms}$ after the first contact 
between the two neutron stars
the amount of mass ablated by the wind is $\approx 2\times 10^{-3}\ M_{\odot}$.
At this stage, the winds have not yet reached a
complete steady state and they could still be evolved further \citep{Martin2015}. 
Here we use a simulation methodology that is
different from  \citet{Perego2014}: they used
Newtonian hydrodynamics with self-gravity, a
spectral neutrino leakage scheme and a tabulated, 
nuclear EoS, while our simulations are special
relativistic with a point-mass source of gravity, 
no neutrino transport and an adiabatic EoS. Both
simulation methodologies therefore find slightly
different equilibria, with the result that our wind
model is initially slightly out of equilibrium.
One of the major deliverables of  
\citet{Perego2014}'s wind simulation is the spatial distribution 
of the  electron fraction $Y_\mathrm{e}$. We start with exactly
this $Y_\mathrm{e}$ distribution and advect it with the flow
as a passive scalar. Both the dynamics and the electron fraction distribution for this initial configuration are shown in Fig.~\ref{fig:windmaps}.
The pressure normalisation is given by the original wind simulation. The whole system is evolved in the gravitational field of the hyper-massive neutron star, which we approximate as a Newtonian point mass of $M_{\rm{HMNS}}= 2.7\ M_{\odot}$ located at the origin.

\subsubsection{Computational hydrodynamics grid}
\label{sec:setup}

The computational grid is roughly shaped as a box ($-10^5 \leq x \leq 10^5$, $-10^5 \leq y \leq 10^5$, $0 \leq z \leq 1.6 \times 10^5$,  measured in km) that is located
above the equatorial plane (at $z = 0$).
Since the merger can be assumed to be symmetric
about the equatorial plane, we use a reflective
boundary condition at the bottom of the
computational domain and outflow boundaries 
elsewhere.
Around the wind ($r \lesssim 2 \times 10^3$ km) and the rotational axis ($r\lesssim 5 \times 10^2$ km)
we fix the numerical resolution to the highest
refinement level (resolution length $\approx 6$ km),
and we use lower resolutions elsewhere.
Before outflowing matter can reach a
boundary, we re-map the matter configuration
into a larger domain to allow further, unhindered
evolution. The simulations are stopped once the
wind material is expanding roughly homologously (at $t \gtrsim 1$ s). Once this stage is
reached, we can scale the matter homologously
to the larger distances so that it can be straight-forwardly used in  the radiative transfer simulations (see Section~\ref{sec:radsetup}).

\subsubsection{The ambient medium}
\label{sec:ambient}
The evolution of the original wind \citep{Perego2014} has
been performed with a background medium of 
$\rho_{\rm amb}=5\times10^3\ \mbox{g cm}^{-3}$, a 
value set by the bottom value of  the tabulated 
nuclear EoS that is used in those simulations. Since
we are using a simple polytropic equation of state 
this bound does not apply here and we embed the wind
initial data in a steeply decreasing background density
with $\rho(r)= \rho_0 \left( R_0/r\right)^4$, where 
$\rho_0 =10^{-6}\ \mbox{g cm}^{-3}$ and $R_0$ is close 
to the upper wind boundary at $2000\ \mbox{km}$. 
We choose this environment with the purpose of reaching quickly and without discontinuities very low matter densities. This profile is steeper than profiles of stationary winds ($\propto r^{-2}$), and it has no impact on the results as long as its energetic contribution in the system is negligible compared to the one from the wind and the jet.
We set the electron fraction inside the background 
material to an artificially high value of 
$Y_{\rm e}^{\rm amb}=1$ so that it is easily identifiable
throughout the entire simulation. 
 \\

\subsubsection{The jet}
\label{sec:jet_inj}

We assume here that a relativistic jet has already 
formed and has reached  a height of $40\ \rm{km}$ above 
the remnant, and from this point onward
propagates into the wind \citep{Gottlieb2018a, Harrison2018, Mizuta2009, Mizuta2013}. 

We  model the jet as an unmagnetized conical outflow with
an opening angle of $\theta_0 = 5^{\circ}$. We inject it
through inflow boundary conditions close to the grid origin. 
The jet is parameterised by its (total) luminosity
$L_\mathrm{j}$, its initial, $\Gamma_0$, and asymptotic
Lorentz factor $\Gamma_\infty = h_0 \Gamma_0$, where $h_0$
is the initial specific enthalpy. The three components of 
the speed are obtained from the jet geometry and
$\Gamma_0$, while the density is obtained from
\begin{ceqn}     
\begin{equation}
  L_{\rm j} =  \Gamma_{0}^2\rho_{\rm j}h_0c^2\beta_{\rm j}\Sigma_{\rm j},
 \end{equation}
\end{ceqn}
where $\Sigma_{\rm j}$ is the cross-sectional area of the jet at the top of the injection region $z_{0}$.
The pressure in the inlet region is set by the previous parameters 
together with the EoS as:
\begin{ceqn}
\begin{equation}
 p_{\rm j} = \frac{\gamma-1}{\gamma}\rho_{\rm j}(h_0-1).
\end{equation}
\end{ceqn}
As for the ambient medium, we set the electron fraction
within the jet to $Y_{\rm e,j}=1$ to keep it easily
recognisable at later times. The jet injection is 
kept at full power from the beginning of the simulation. 

The jet is injected from the beginning of the simulation.
Because of the very specific initial conditions around the launching region a way to recognise if the jet manages to propagate is required. To do so we choose to set a minimum Lorentz factor $\Gamma =2$  and a height above the origin.
Once the head has reached that height we recognise the jet as ``launched", and the 
jet is pushed further for a time of  
$\Delta t_\mathrm{inj} = 100$ ms. After that time 
the luminosity decays exponentially.

We run three simulations: two with jets of different luminosities ($L_\mathrm{j} = 10^{49}$ erg s$^{-1}$, \texttt{Jet49}, and $L_\mathrm{j} = 10^{51}$ erg s$^{-1}$, \texttt{Jet51}) and, as a reference case, we evolve
in one simulation (\texttt{Wind}) only the wind without
injecting  an additional jet. Our chosen values for
$L_\mathrm{j}$, $\Gamma_0$ and $\Gamma_\infty$ are
representative for low- and high-luminosity jets in GRBs \citep{Fong2015}. All our jet parameters are listed in Table~\ref{tab:initial_par}.

\begin{table}
	\centering
	\caption{Jet parameters for the current simulations: initial opening angle $\theta_0$, height of the injection region $z_0$, effective jet duration (from launching) $\Delta t_{\rm inj}$, number of cells covering the injection region in the x direction $N_{\rm inj,x}$, initial luminosity $L_{\rm j}$ and initial $\Gamma_0$ and asymptotic Lorentz factor $\Gamma_{\infty}.$}
	\label{tab:initial_par}
	\begin{tabular}{l|c}
		\hline
		\ & Geometry \\
		\hline
		$\theta_{0}$ & $5^{\circ}$ \\
		$z_{0}$ [km] & $40$  \\
		$\Delta t_{\rm inj}$ [ms] & $100$ \\
		$N_{\rm inj,x}$ & 8 \\
		\hline
		\ & Physical\ parameters \\
		\hline
		$L_{\rm j}$[$\rm{erg\ s^{-1}}$] & $10^{49}$, $10^{51}$  \\
		$\Gamma_0$ & $10$ \\
		$\Gamma_{\infty}$ & $200$ \\
		\hline
	\end{tabular}
\end{table}

\begin{figure*}
 \centering
  \subfloat{
	\begin{minipage}{
	   0.48\textwidth}
	   \centering
	   \includegraphics[width=1\textwidth]{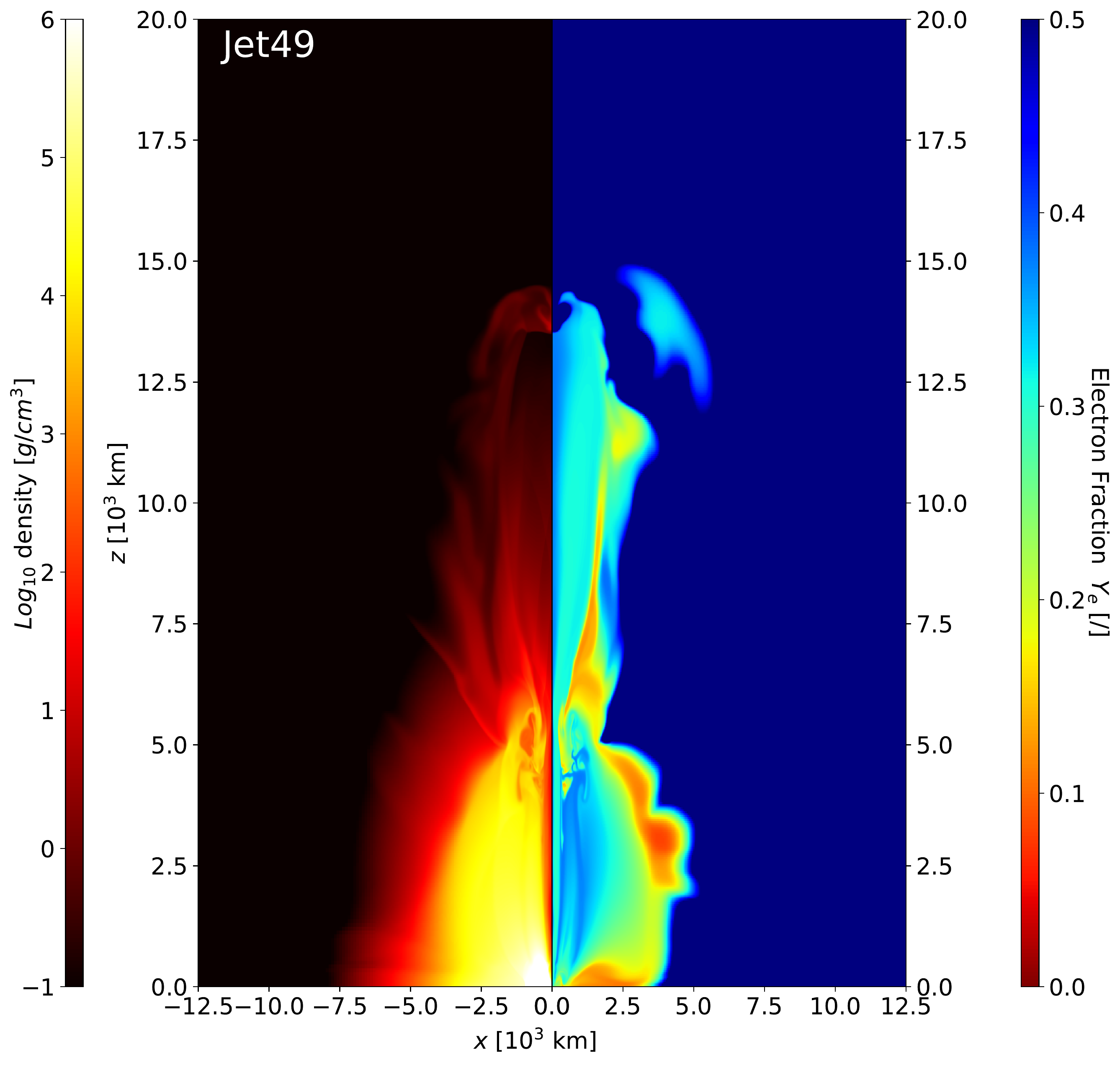}
	\end{minipage}}
  \quad 
  \subfloat{
	\begin{minipage}{
	   0.49\textwidth}
	   \centering
	   \includegraphics[width=1\textwidth]{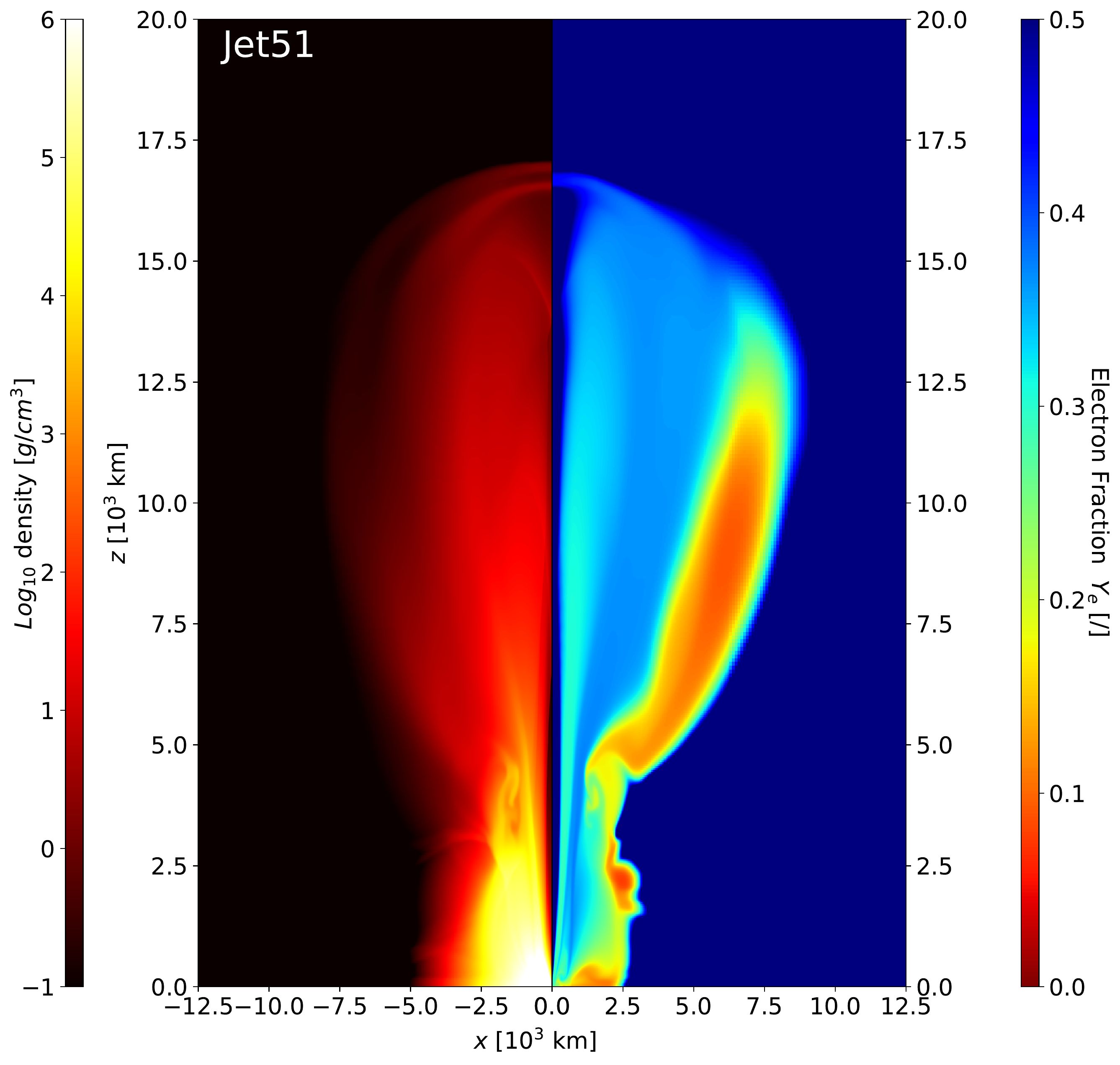}
	\end{minipage}}
    \caption{Vertical slices (the $y=0$ plane) for the two different jet models: \texttt{Jet49} on the left and \texttt{Jet51} on the right. Both panels show the rest-mass density distribution on a logarithmic scale (left-hand side) and electron fraction distribution (right-hand side). In the $Y_{\rm e}$ map colours from red to yellow mark the lanthanide-rich ejecta, while the light-blue one represent the lanthanide-poor.
    Since the jet launch occurs at different times for
    the two cases we show both cases at 60 ms after
    jet launch (roughly 65 (\texttt{Jet51}) and 115 (\texttt{Jet49}) ms from the beginning of the simulation, corresponding to 170 and 220 ms after the first contact).
    }
    \label{fig:jetmaps}
\end{figure*}

\begin{figure*}
	\centering
	\includegraphics[width=0.8\textwidth]{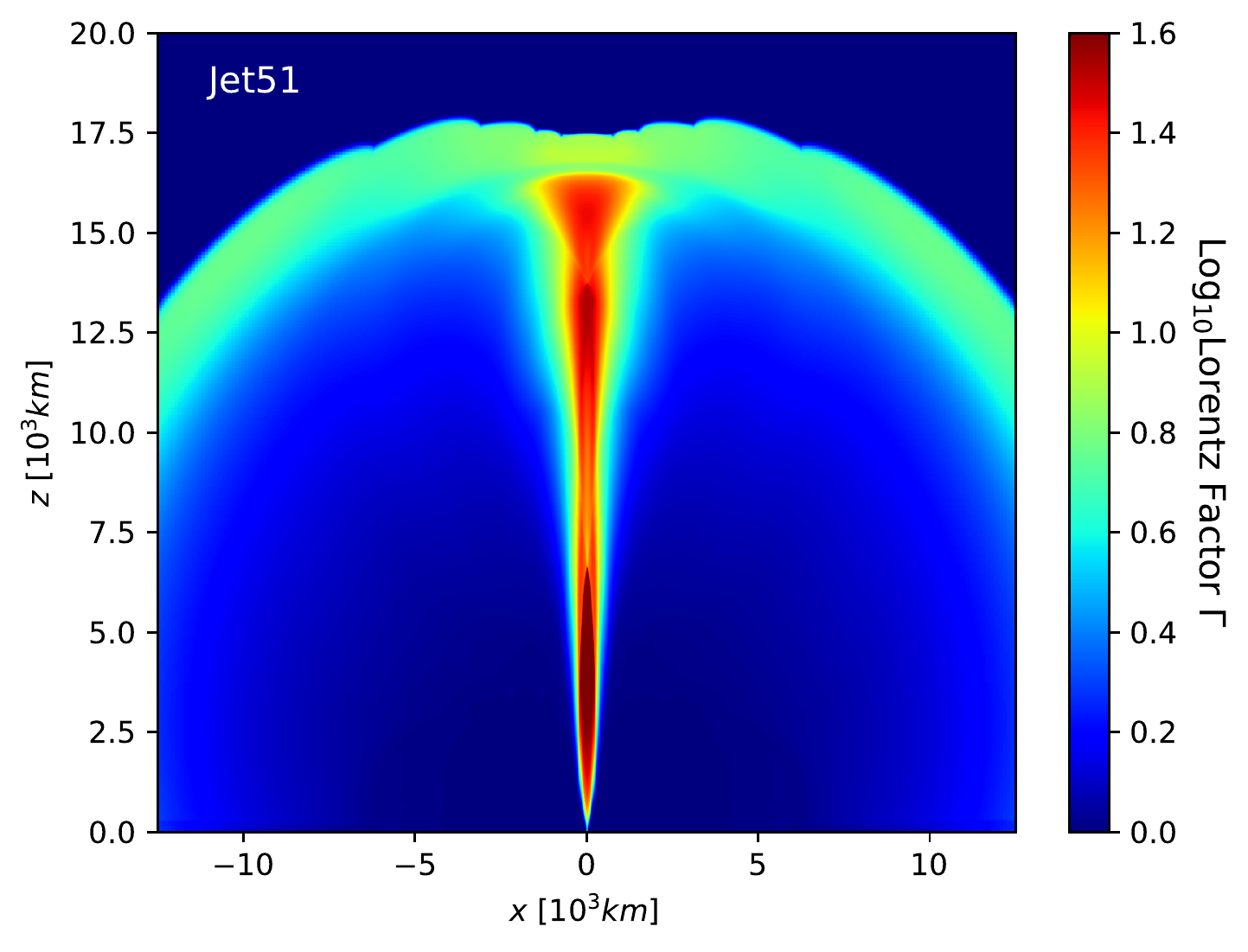}
    \caption{  Vertical slice ($y=0$ plane) of the log-scaled distribution of Lorentz factor $\Gamma$ for the jet model \texttt{Jet51} from the same snapshot of the Fig.~\ref{fig:jetmaps} (right). The jet undergoes a strong first collimation shock and stays collimated after breaking out from the ejecta. (The leading shock is an artifact from our chosen density and pressure gradients in the ambient medium, but carries essentially no mass and energy and therefore has no impact on the simulation.)
    }
    \label{fig:lfacmaps}
\end{figure*}

\subsection{Radiative transfer setup}
\label{sec:radsetup}

Radiative transfer simulations are performed for the models
\texttt{Wind}, \texttt{Jet49} and \texttt{Jet51}
introduced in Section~\ref{sec:jet_inj}. In particular, 
the grid domain is restricted for the two jet models 
to be the same as in the \texttt{Wind} model, with a
maximum velocity of $\sim0.35$\,c (maximum spatial
coordinate of $\sim10^5$\,km at 1\,s after the
merger)\footnote{This cut in the jet models does 
not affect the final observables since the densities 
and corresponding opacities of the removed material 
are negligible.}. Each model grid is symmetrised 
about the orbital plane and downgraded to a 
uniform Cartesian grid with 128$^{3}$ cells.
As mentioned in the introduction, the potentially dominant
contributions to the ejecta come from rather slow matter parts
that are unbound on a secular time scale from the accretion
torus. Their simulation is computationally extremely
expensive and to date the properties of such outflows
are still not entirely settled. There is agreement
that about 40\% of the initial torus mass becomes
unbound \citep{just15,siegel17a,siegel18,miller2019,fernandez19}, but there is no consensus yet, whether the outflows are lanthanide-rich \citep{siegel18} or lanthanide-poor \citep{miller19}. We therefore model this ejecta component as a spherical mass distribution with a density profile as in equation~(10) of \cite{Wollaeger2018} and total mass of
0.072 M$_\odot$ (corresponding to 40\% of our initial
torus mass). We release this mass 1 s
after merger with velocities distributed between 0.03 to 0.1\,c. 
Since the composition  is
currently unsettled, we perform each time two simulations
once assuming a lanthanide-poor and once a lanthanide-rich
composition. We note that the disk ejecta contribute to
$\sim$\,99.5, 99.5 and 98.8\,\% of the total mass for the
\texttt{Wind}, \texttt{Jet49} and \texttt{Jet51} model,
respectively. The final grid is expanded to $t_i=0.1$\,d
and densities are scaled as $\propto t^{-3}$ according to
homologous expansion.

The simulations with lanthanide-poor (-rich) compositions are carried out with $N_{\rm ph}=2\times10^7$ ($1.5\times10^7$) photons. 
Viewing-angle dependent SEDs are computed from $t_i=0.1$ to $t_f=20$\,d after the merger and used to create synthetic $ugriz$ light curves from different orientations. 
All the models investigated are sufficiently close to axial symmetry about the jet axis $z$ and symmetric about the orbital plane ($xy$) by construction. Therefore, we restrict our analysis to orientations in the $xz$ plane ($\phi_{\rm obs}=0^\circ$) and extract light curves for $N_{\rm obs}=11$ observers from the jet axis ($\theta_{\rm obs}=0^\circ$) to the orbital plane ($\theta_{\rm obs}=90^\circ$) equally spaced in cosine ($\Delta\cos\theta_{\rm obs}=0.1$). Different opacities are assumed \citep{Bulla2019b}  depending on whether the composition of the ejecta is lanthanide-poor ($Y_{\rm e}\geq0.25$) or lanthanide-rich ($Y_{\rm e}<0.25$). The nuclear heating rates are taken from equation~(4) of \citet{Korobkin2012} while thermalization efficiencies 
come from \citet{Barnes2016}.

\section{Results}
\label{sec:results}
\subsection{Hydrodynamic evolution}
The hydrodynamic evolution is broadly consistent with what is expected from the literature 
\citep{Zhang2003, Mizuta2009, Bromberg2011, Mizuta2013, Nagakura2014, Murguia-berthier2014, Gottlieb2018a, Duffell2018} 
and the aforementioned theoretical framework provides all that is needed to understand the main features of jet propagation.

Our initial conditions as obtained from the 
neutrino-driven wind simulations of \cite{Perego2014} 
have a peculiar $Y_e$ distribution: their leading 
edge is made of very low $Y_{\rm e}$ material  
that could potentially block emission as a ``lanthanide
curtain" \citep{kasen15,Wollaeger2018}. It is interesting to see what happens to this effective 
``high opacity skin" during the further hydrodynamic 
evolution and what its impact is on the electromagnetic emission.

At the beginning of our simulations, the matter contains expanding wings that flow outwards at angles of 
$\sim 30$ degrees from the rotation axis (see Fig.~\ref{fig:windmaps}). Furthermore, matter without centrifugal support is falling towards the central neutron star along the rotational axis. The downfall compresses the gas in the 
injection region, which initially hampers a jet launch. 
A successful launch requires the total jet momentum
flux at the jet head to overcome the gas pressure and ram pressures, 
$P_{\rm j, ram} > max(P_{\rm gas}, P_{\rm gas, ram})$. 
In nature, this condition may be reached once enough
energy has been deposited to push the gas aside or
the central neutron star collapses into a black hole 
so that the region around the rotation axis is cleaned from polluting baryons \citep{Macfadyen1999}. In our simulations, the jets
only manage to overcome surrounding pressure after times
of $t_{\rm inj}\approx 5\ \rm{ms}$ in the \texttt{Jet51}
model and $t_{\rm inj}\approx 50\ \rm{ms}$ in the 
\texttt{Jet49} model, approximatively $100$ and $150$ ms from the first contact respectively. These numbers are broadly consistent
with the estimates of \cite{Beniamini2020} for jet
formation time scales.

As shown in Fig.~\ref{fig:jetmaps}, both jet models
fill pressurized cocoons, wider and more energetic 
in the high-luminosity case. Despite their (two orders of
magnitude) different jet powers, there are successful
breakouts in both cases. 
We follow the approach of \cite{Duffell2018} to estimate
which minimum energy is needed for a successful jet break
out. Using our simulation results\footnote{We use the coefficient calibrated on simulations in \cite{Duffell2018}, there named $\kappa\approx 0.05$.}, we find a threshold value
of $E_{\rm crit}\approx 5\times 10^{45}\ \rm{erg}$,
i.e. even for our low luminosity case with 
$E_{\rm jet}=10^{48}\ \rm{erg} \gg E_{\rm crit}$, a successful breakout is expected.
When breaking out, the jets push aside the high opacity
skin and make the inner, lower opacity regions more accessible to potential observations. This will be
discussed in more detail in the next section.

In passing we also want to briefly mention jet collimation
by the surrounding high-pressure cocoon.
Both jets start with a constant opening angle, and their
initial conical shape gets roughly cylindrical during
propagation, after the collimation from the surroundings. 
To estimate whether collimation is expected or not, one
can resort to the dimensionless jet luminosity parameter \citep{Bromberg2011}:
\begin{ceqn} 
\begin{equation}
    \Tilde{L}\equiv \frac{\rho_{\rm j}h_{\rm j}\Gamma_{\rm j}^2}{\rho_a}
    .
\end{equation}
\end{ceqn}

For the \texttt{Jet51} model we find $\tilde{L}\lesssim 5 \lesssim \theta_{0}^{-4/3}$ and therefore at least one collimation shock is therefore expected, consistent with
 the Lorentz factor map in Fig.~\ref{fig:lfacmaps}. 
 The jet is significantly decelerated as a result of 
 a very strong first collimation shock. After the 
 breakout, the jet remains at first collimated, but 
 starts later to spread sideways, consistent with the
 results by \cite{Mizuta2013} and \cite{Nagakura2014}. In model \texttt{Jet49}, collimation is even stronger and 
 it happens earlier and closer to the injection region
 \citep{Matzner2003, Bromberg2011, Mizuta2013, Harrison2018}. 
 It changes quickly from conical to cylindrical and stays very
 narrow until breakout, experiencing multiple 
 recollimation shocks. In this model, a shocked jet 
 is observed spreading immediately after the 
 breakout and returning to a quasi-conical 
 configuration. Further studies of these 
 features are left for more dedicated future
 investigations.

\begin{figure*}
    \centering
	\includegraphics[width=0.98\textwidth]{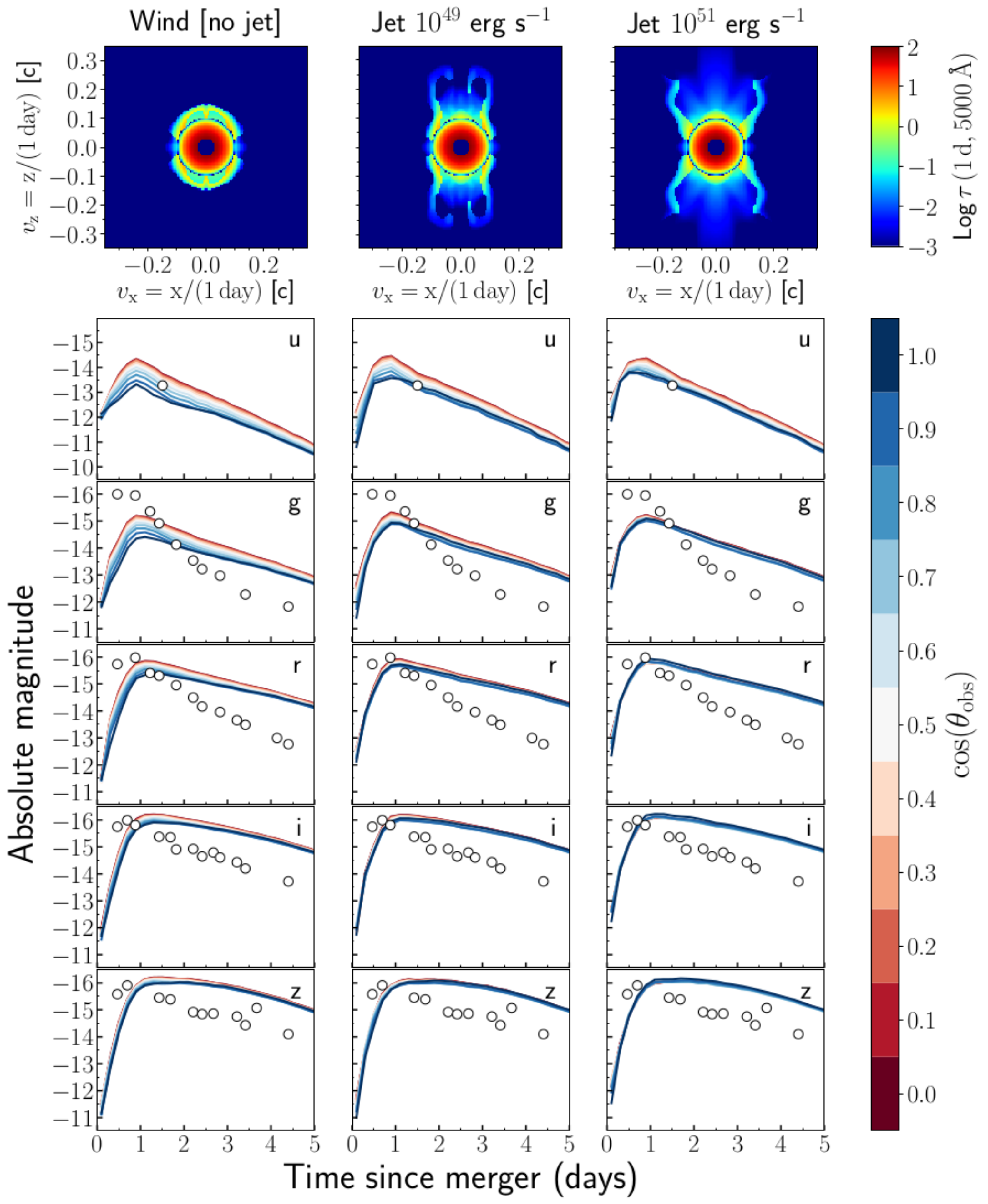}
    \caption{Broad-band light curves for the \texttt{Wind} (left), \texttt{Jet49} (middle) and \texttt{Jet51} (right) models with a lanthanide-poor disk added post-processing between $0.03\,$c and $0.1\,$c. \textit{Top panels}: opacity at 1\,day and 5000\,\AA{} ($\sim$~$g$-band) from \citet[][see their fig. 2]{Bulla2019b}: $\kappa=0.5$\,cm$^2$\,g$^{-1}$ for lanthanide-poor regions with $Y_{\rm e}\geq0.25$ and $\kappa=10$\,cm$^2$\,g$^{-1}$ for lanthanide-rich regions with $Y_{\rm e}<0.25$. \textit{Bottom panels}: $ugriz$ (from top to bottom) light curves for different viewing angles, going from the jet axis ($x=0$, dark blue) to the orbital plane ($z=0$, dark red).
    Broad-band photometry of the GW\,170817 MKN AT\,2017gfo is shown with white circles for comparison. }
    \label{fig:lc_lf}
\end{figure*}

\begin{figure*}
    \centering
	\includegraphics[width=0.98\textwidth]{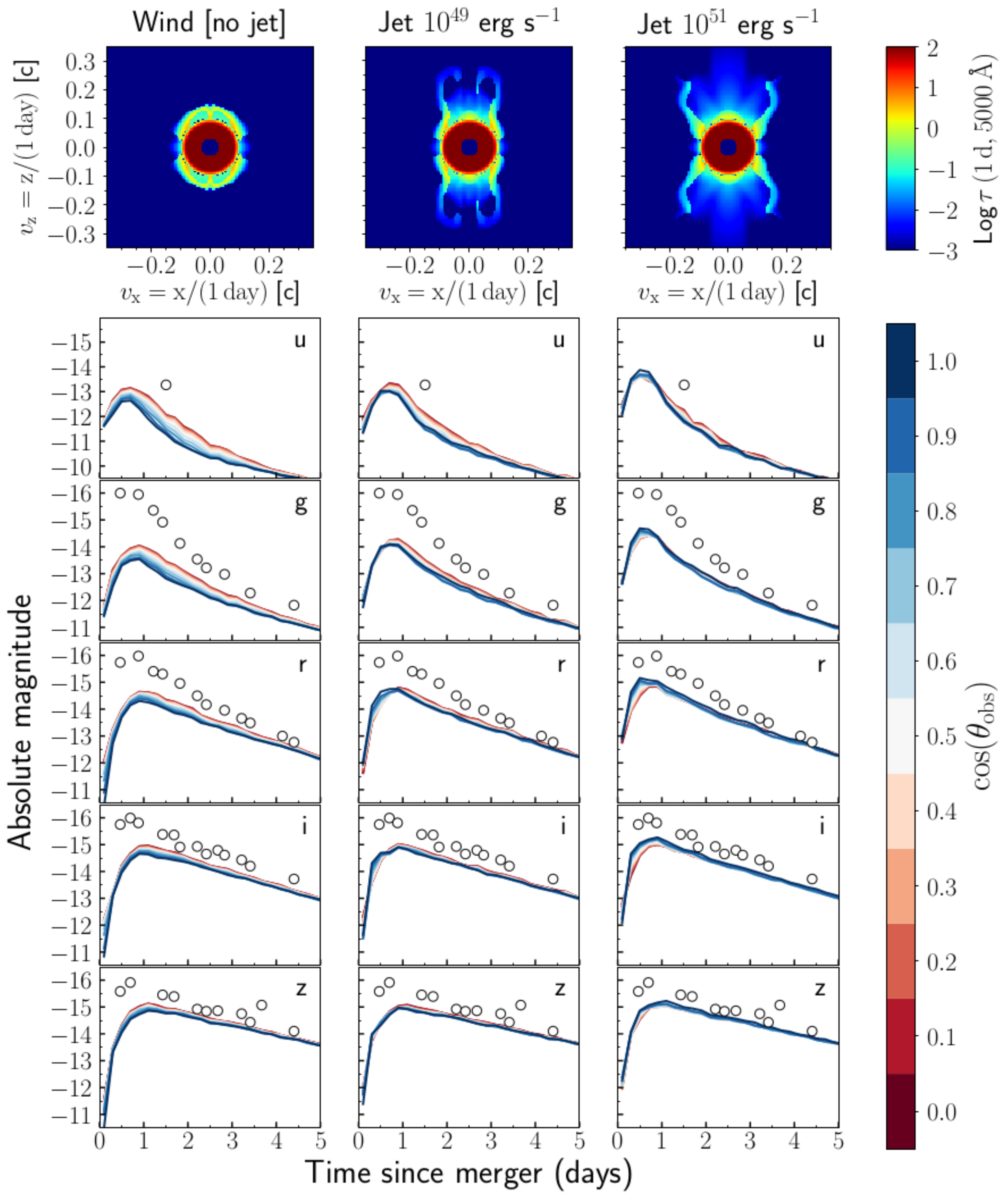}
    \caption{Same as Fig.~\ref{fig:lc_lf} but assuming a lanthanide-rich composition for the inner disk between $0.03\,$c and $0.1\,$c.}
    \label{fig:lc_lr}
\end{figure*}

\begin{figure*}
    \centering
	\includegraphics[width=0.49\textwidth]{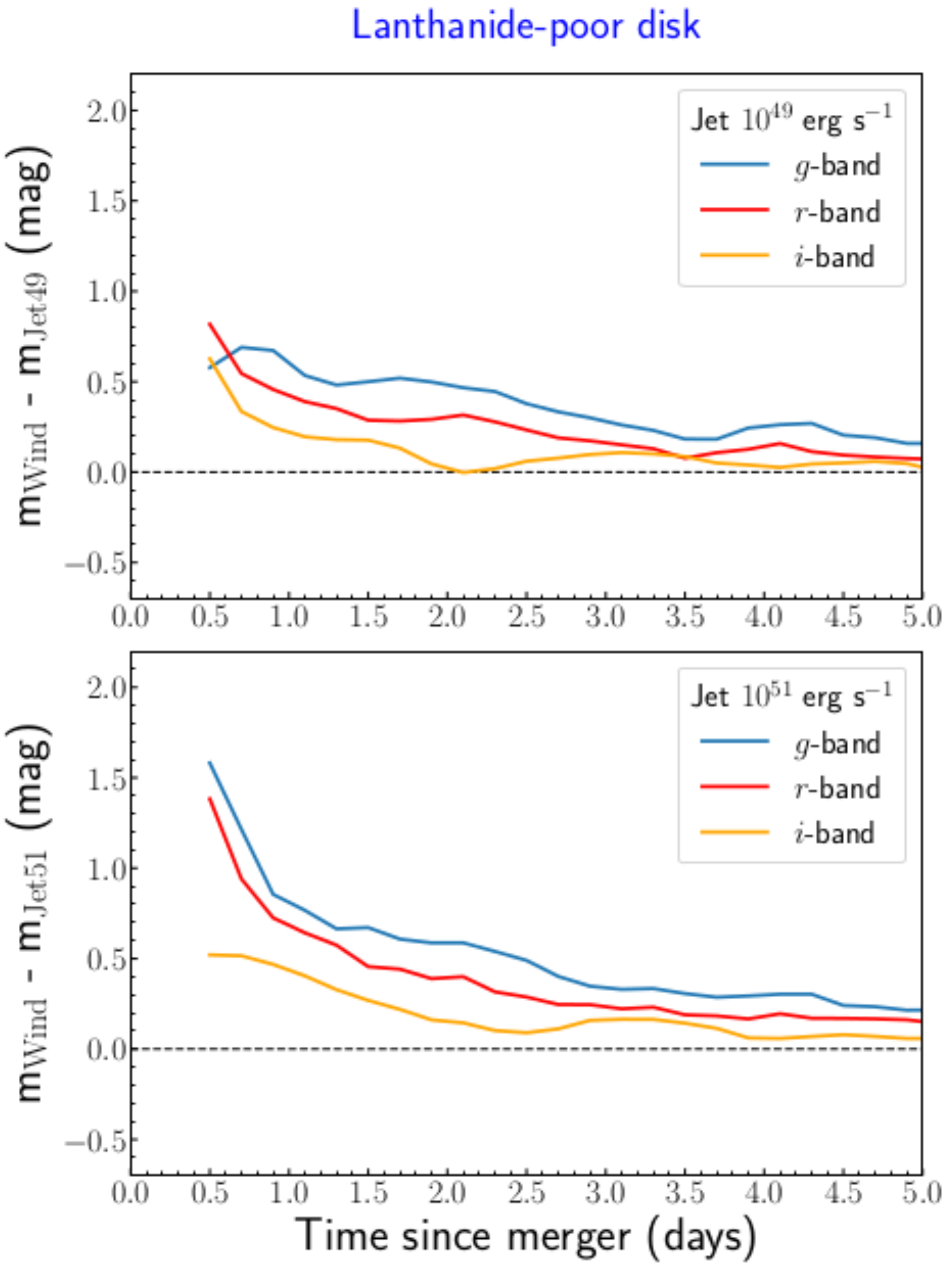}
	\includegraphics[width=0.49\textwidth]{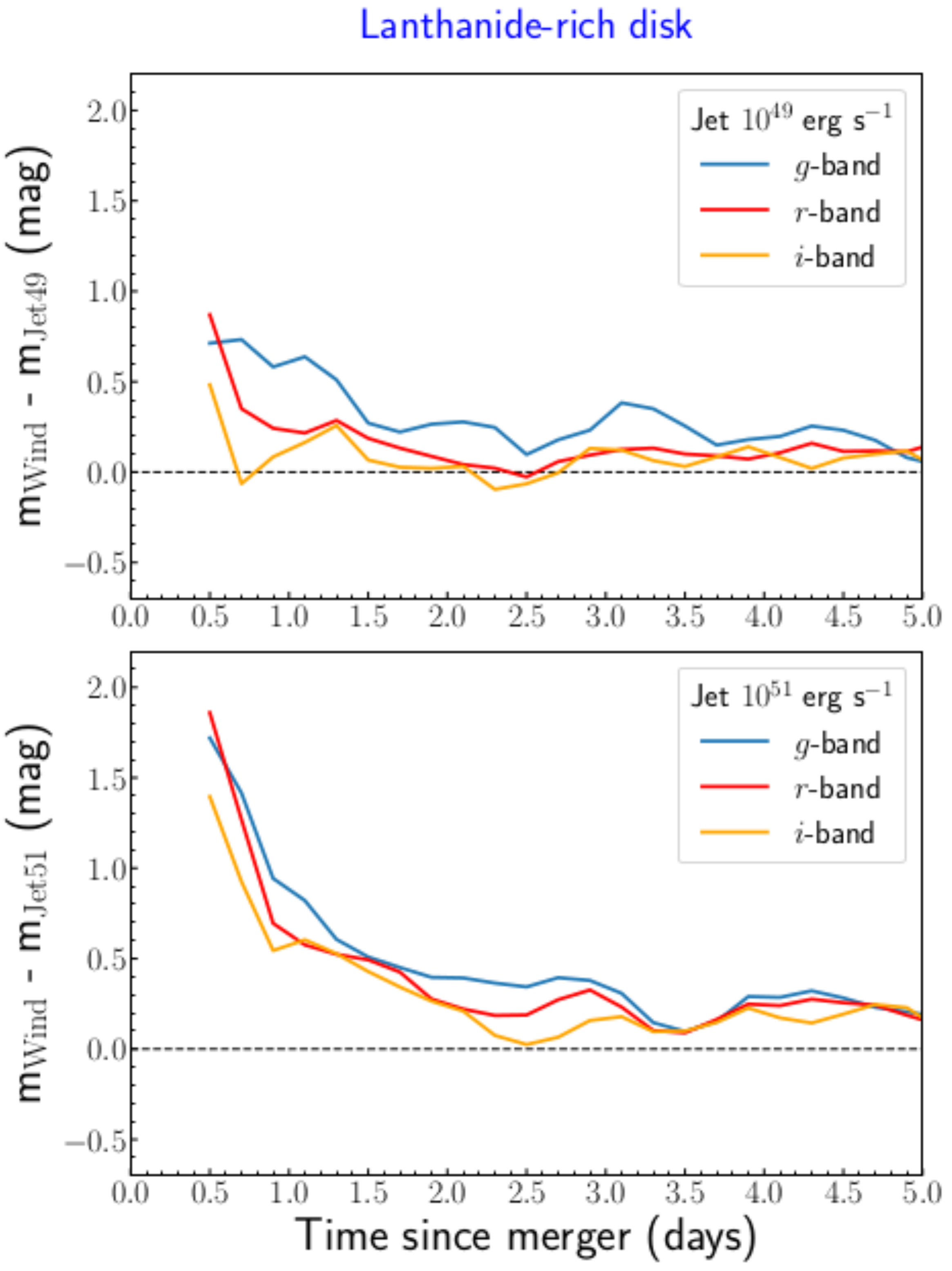}
	\caption{Magnitude change in the $gri$ filters induced by the jet close to the rotational axis ($\cos\Theta_{\rm obs}\geq0.9$, dark blue lines in Figs.~\ref{fig:lc_lf}-\ref{fig:lc_lr}). Predictions are shown as a function of time since merger, for models with lanthanide-poor (left) and lanthanide-rich (right) disks. Top panels refer to the \texttt{Jet49} models, while bottom panels to the \texttt{Jet51} models. The increase in brightness due to the jet punching is stronger for more powerful jets, bluer filters and lanthanide-rich disks.}
    \label{fig:deltam}
\end{figure*}

\subsection{Radiative signatures}
In this Section, we present broad-band light curves extracted with the radiative transfer code \textsc{possis} for the \texttt{Wind}, \texttt{Jet49} and \texttt{Jet51} models. 
In Section~\ref{sec:lpoor} we
present the simulations with a lanthanide-poor disk while those 
with a lanthanide-rich disk are discussed in Section~\ref{sec:lrich}.

\subsubsection{Lanthanide-poor disk}
\label{sec:lpoor}
Fig.~\ref{fig:lc_lf} shows $ugriz$ light curves predicted for the wind and jet models assuming a lanthanide-poor composition for the  disk ejecta. In general, we do see variations in both the brightness and viewing-angle dependence between the \texttt{Wind} model (left panels) and the two models with jets (middle and right panels). As we describe in the following, these differences can be understood by a close inspection of the opacity maps, which are given in the top panels of Fig.~\ref{fig:lc_lf} for each model.

In the \texttt{Wind} model, material at high latitudes (close to the $z$ axis) is characterised by lower $Y_{\rm e}$ and thus higher opacities from lanthanides compared to material in the orbital plane. As a consequence, radiation can escape more easily in the orbital plane than through this low-$Y_{\rm e}$ ``curtain'' close to the rotational axis. This effect leads to a clear viewing-angle dependence in the light curves, with orientations close to the orbital plane ($\cos\theta_{\rm obs}=0$, dark red light curves) associated to brighter MKNe compared to those along the $z$ axis ($\cos\theta_{\rm obs}=1$, dark blue light curves). Since bound-bound line opacities become larger when moving to shorter wavelengths \citep{Bulla2019b}, the viewing-angle effect is  stronger in bluer filters. For instance, the peak brightness in the $u$ band is $\sim1$~mag brighter for an observer in the orbital plane compared to one on the $z$ axis. This difference is, instead, small ($\Delta m\lesssim$~0.3 mag) in the redder $iz$ filters.

In the \texttt{Jet49} and \texttt{Jet51} models, the low-$Y_{\rm e}$ ``curtain'' close to the rotation axis in the \texttt{Wind} model is ``punched'' away and decreased in density by the jet (see ``arm''-like structures in the top panels of Fig.~\ref{fig:lc_lf}). This effectively reduces the opacities of the material surrounding the disk ejecta, leading to a clear imprint of the jet on the final observables. Since the ``punch-away'' mechanism is restricted to regions at high latitudes, an increase in brightness is found for orientations close to the rotation axis (blue lines). As highlighted in the left panels of Fig.~\ref{fig:deltam}, this increase is restricted to the first $\sim$ 3 days after the merger while negligible at later epochs when ejecta become optically thin outside the  spherical viscous ejecta. While the jet makes MKN brighter at high latitudes, the increase in brightness is nearly absent for orientations in the orbital plane (red lines), therefore decreasing the viewing-angle dependence seen in the \texttt{Wind} model. It is worth noting that this would correspond to a jet \textit{introducing} a viewing-angle dependence for an initial spherical distribution of the winds. In addition, we find that the increase in brightness and decrease in viewing-angle dependence is stronger moving from redder to bluer filters due to line opacities being larger at shorter compared to longer wavelengths. Hence, MKNe are made bluer by the jet in the first $\sim$~3 days after merger, with $\Delta(g-r)\sim0.1$ mag and $\Delta(g-i)\sim0.3$ mag (see Fig.~\ref{fig:deltam}). Fig.~\ref{fig:deltam} also highlights how the ``punch-away'' mechanism and the corresponding impact on the observables are stronger for the \texttt{Jet51} model characterised by a more powerful jet. For an observer along the jet axis (polar view), the increase in the $gr$ filters in the first day after the merger is of $\sim1-1.5$~mag for the \texttt{Jet51} model while of $\sim0.5-1$~mag for the \texttt{Jet49} model.

The bottom panels of Fig.~\ref{fig:lc_lf} include $ugriz$ light curves for the MKN associated with GW\,170817, AT\,2017gfo \citep{Andreoni2017,Arcavi2017,Cowperthwaite2017,Kasliwal2017Sc,Pian2017,Smartt2017,Tanvir2017,Valenti2017, Evans2017}. While the overall peak brightness predicted by all the models is comparable to that observed in AT\,2017gfo, we note a clear difference in the light curve evolution. In particular, models are systematically fainter than the data in the first day after the merger (especially in $gri$ optical filters) and systematically brighter afterwards. 

\subsubsection{Lanthanide-rich disk}
\label{sec:lrich}

Fig.~\ref{fig:lc_lr} shows $ugriz$ light curves predicted for the wind and jet models assuming a lanthanide-rich composition for the  viscous ejecta. The behaviours identified in Section~\ref{sec:lpoor} for the lanthanide-poor case are visible in the lanthanide-rich case as well. Namely, we find a modest viewing-angle dependence in the bluer filters of the \texttt{Wind}
model, while an increase in brightness and decrease in viewing-angle dependence when a jet is launched in the system. As shown in Fig.~\ref{fig:deltam}, the increase in $gri$ magnitudes for a polar viewing angle are relatively similar between jet models with a lanthanide-poor and lanthanide-rich disk. The increase in brightness is limited to the first $\sim3$ days and of the order of $\sim0.5$ mag in the \texttt{Jet49} and $\sim0.5-1.5$ mag in the \texttt{Jet51} models in the first day.

Although the increase in brightness due to the presence of a jet is similar regardless of the disk composition, we do see a clear difference between the two sets of models. In particular, a lanthanide-rich composition of the disk ejecta is associated with higher opacities in the inner regions of the ejecta. The predicted MKN light curves are therefore much fainter than in the lanthanide-poor case, in all filters and especially after peak. As a consequence, the MKN light curves predicted by the \texttt{Wind}, \texttt{Jet49} and \texttt{Jet51} are fainter than those observed in AT\,2017gfo, especially at early epochs. Although our models are not tailored to provide good fits to AT\,2017gfo, we do note that the \texttt{Jet51} model with a lanthanide-rich disk produces the closest agreement to the observed MKN. In particular, the predicted light curves peak at $\sim0.5-1$ mag lower magnitudes in all filters and have very similar decays to those observed (cf with the lanthanide-poor case, Fig.~\ref{fig:lc_lf}).

\section{Discussion and conclusions}
\label{sec:summary}
We have performed a set of 3D special-relativistic
simulations where we study the interaction of jets
with the neutrino-driven winds from a neutron star merger
remnant. Our particular focus is on the question how this 
interaction impacts on the resulting MKN light curves. 
The hydrodynamic simulations have been performed with the
special relativistic AMR hydrodynamics code \textsc{amun} and
the radiative signatures are subsequently extracted with
the Monte-Carlo radiative transfer code \textsc{possis}.
\par 
The electron fraction $Y_e$  plays a crucial role for neutron star 
merger ejecta: it determines the nucleosynthesis, the nuclear 
heating rates, the opacities within the ejected matter and 
therefore the properties of the resulting electromagnetic transients.
Despite of this, most previous studies started from highly idealised
and mostly guessed initial conditions. In our study we use a 
realistic post-merger environment based on the neutrino-driven wind
simulations of \cite{Perego2014} as initial conditions. 
These winds provide us with information about the dynamics and
in particular with a peculiar and non-trivial $Y_{\rm e}$ 
distribution. There are downflows along the original
binary rotation axis towards the central neutron star which
create a high-pressure environment that initially hampers 
jet formation so that we need to trigger the jets in our numerical
models for a while before they are finally launched.
\par We explore two different types of jet models with 
$L_\mathrm{j} = 10^{49}$ erg s$^{-1}$ (\texttt{Jet49}) and
$L_\mathrm{j} = 10^{51}$ erg s$^{-1}$ (\texttt{Jet51}) 
representative of typical low and high luminosity GRBs. As a reference we have also 
run a model where only
the ejecta are evolved and no jet is injected (model \texttt{Wind}).
The dynamical evolution of our jet models is generally in good agreement with the
expectations from the literature \citep{Zhang2003, Mizuta2009, Bromberg2011, Mizuta2013, Nagakura2014, Murguia-berthier2014, Gottlieb2018a, Duffell2018,Murguia20}. Both our  models show 
recollimation shocks, but especially the features in the 
\texttt{Jet49} model suggest a strong collimation 
which might potentially leave an imprint in the early 
gamma-ray signal.

\par We use the radiative transfer code \textsc{possis} \citep{Bulla2019b} to predict viewing-angle dependent MKN light curves for the \texttt{Wind}, \texttt{Jet49} and \texttt{Jet51} models. The radiative transfer simulations are performed on 
a matter background for which we use the results from our relativistic hydrodynamics simulations enhanced by a matter component that models the secular ejecta from a central
accretion torus.
In summary, we find the following:

\begin{itemize}[wide=0pt, leftmargin=\dimexpr\labelwidth + 2\labelsep\relax]
\setlength\itemsep{0.7em}
\item The models with no jet (\texttt{Wind} models) show some viewing-angle dependence in the predicted light curves, being fainter for orientations close to the rotation axis. This effect, caused by low-$Y_{\rm e}$/high opacity material (``curtain'') near the jet axis, is stronger when moving to shorter wavelengths (e.g. $\Delta m\sim1$ mag in the $g$-band while $\Delta m\lesssim0.3$ mag in the $i-$band). 
\item In the \texttt{Jet49} and \texttt{Jet51} models, the low-$Y_{\rm e}$/high opacity ``curtain'' close to the axis is ``punched-away'' and decreased in density by the jet. Compared to the \texttt{Wind} model, radiation can therefore escape more easily along the jet axis, an effect that is stronger in the near-UV and at short optical wavelengths where line opacities are typically higher;
\item As a consequence, the presence of a jet makes MKNe \textit{brighter} and \textit{bluer} in the first $\sim3$ days after the merger for observer orientations close to the jet axis. The presence of a strong jet seems to erase the viewing angle dependence seen in the \texttt{Wind} model, hence making the emission appear more isotropic.
\item The increase in brightness is stronger for a more powerful jet, with the \texttt{Jet49} and \texttt{Jet51} models being $\sim0.7$ and 1.5 mag brighter than the \texttt{Wind} model in the first day after the merger in $g$ and $r$ filters;
\item  Since to date there is no consensus about the 
composition of secular ejecta from the inner accretion torus, we have each time performed a simulation with a
lanthanide-free and a lanthanide-rich composition.
Although we had no intention to specifically model 
the first detected neutron star merger GW\,170817/AT\,2017gfo, it is worth stating that 
the models with a lanthanide-rich inner disk are 
fainter and in better agreement with the observations 
than models with a lanthanide-poor disk.
\end{itemize}

The jet, with its quasi-radial velocity profile and high pressure, tends to transport material away from the axis, which decreases the on-axis opacity. The jet also gives energy to the material that it interacts with, so that the material reaches a higher ballistic speed, which translates to a lower density at the time of the MKN emission, and therefore a lower opacity.
However, in this work we have not considered the details of jet formation. There exists a possibility that the process of jet formation could drag with it large amounts of high opacity material that would otherwise stay bound, effectively forming an additional merger mass-loss channel. The net effect of such mass-loss on the MKN emission is not known. This process could be studied through jet formation simulations, and is beyond the scope of this paper.

\par Clearly, these findings depend on the presence of a thin low-$Y_e$ and
high-opacity ``skin layer" ($\sim 10^{-5}$ M$_\odot$) in our initial data.
3D neutrino-hydrodynamic simulations of neutron star mergers, of their
remnants and emerging winds are major computational challenges of
contemporary astrophysics and today's results, where comparable,
do not (yet) agree on all aspects. Therefore, it is justified to
ask how real this lanthanide curtain is.
\par The original neutrino-driven wind simulations of \cite{Perego2014} were obtained with
the \textsc{FISH} code \citep{kaeppeli09} enhanced by an advanced leakage scheme \citep{Perego16} that accounts for neutrino absorption in optically thin conditions. As initial
condition, served merger remnant configurations \citep{price06} obtained with the
SPH code \textsc{MAGMA} \citep{rosswog07c}. These latter simulations accounted for weak interactions and in particular for $Y_e$-changes due to electron-/positron  captures, but they did not include neutrino absorption \citep{Rosswog2003a}.
Nevertheless we believe that these simulations provide a fair representation of what happens in nature to $Y_e$ inside the remnant in the early post-merger evolution. 
The initial torus is formed from matter coming from the outer core of the merging neutron stars. The disc formation timescale is expected to be smaller than the timescale over which weak interactions operate inside the disc. Thus, weak interactions hardly have time to substantially change $Y_e$, which stays close to its initial value, i.e. $Y_e \lesssim 0.1$ \citep{Perego2019}.

As the accretion process onto the massive neutron star continues over $\sim 100$ ms, neutrinos emitted from the central remnant and inner torus regions get continuously
absorbed by matter lying further out. They change the electron fraction and deposit their energy there, thus driving a wind with increased $Y_e$. Inside the wind, a spatial gradient in the electron fraction is expected: material that has first reached large distances from the remnant, such as the outer torus layers, has smaller chances
to absorb neutrinos and is therefore likely closer to its original, low $Y_e$ value. This low-$Y_e$ matter is pushed outwards by deeper layers that have
captured neutrinos and therefore forms the leading edge of the wind.
\par While we think that there are good physical reasons for its presence, we cannot
safely exclude a numerical origin of this thin lanthanide curtain or its presence only for a subset of cases. We are
not aware of other studies that have discussed this layer, but this is not
surprising given that a) only few studies include the relevant neutrino physics and reach comparable time scales ($\sim 100$ ms), b) the layer is at the leading edge of the ejecta which first exits the outer boundaries of
the computational domain in a Eulerian simulation and may therefore easily go unnoticed and c) it may only contain a small amount of mass ($\sim$ few $10^{-5}$ M$_\odot$ in our case), hard to track through Lagrangian tracers or fix mesh refinements. But even such a small amount
can, as we have demonstrated here, have substantial observational consequences.

We close by noting that we did not attempt to include 
the very neutron rich ($Y_e \sim 0.05)$, first ejected
tidal dynamical ejecta \citep{rosswog99,bauswein13,hotokezaka13,rosswog13b,radice18a}
which cover mostly the binary orbital plane. This exploration is 
left for future work. While most of the presented results
are of qualitative nature, they clearly illustrate that seemingly
``small details" can have substantial impacts on observable
signatures. As a corollary, this implies that we have to expect
a large variety of electromagnetic transients after the 
merger of two neutron stars.

\section*{Acknowledgements}
This work has been supported by the Swedish Research 
Council (VR) under grant number 2016- 03657\_3, by 
the Swedish National Space Board under grant number 
Dnr. 107/16, the research environment grant 
``Gravitational Radiation and Electromagnetic Astrophysical
Transients (GREAT)" funded by the Swedish Research 
council (VR) under Dnr 2016-06012 and by the
Knut and Alice Wallenberg Foundation. We gratefully 
acknowledge support from COST Action CA16104 
``Gravitational waves, black holes and fundamental physics" (GWverse) and from COST Action CA16214 ``The multi-messenger physics and astrophysics of neutron stars" (PHAROS).\\
The simulations were performed on resources
provided by the Swedish National Infrastructure for
Computing (SNIC) at Beskow, Tetralith and Kebnekaise
and on the resources provided by the North-German
Supercomputing Alliance (HLRN).\\
G.K. acknowledges support from CNPq (no. 304891/2016-9).

\section*{Data Availability}
 
The data that support the observational findings of this study (simulated SEDs and broad-band light curves) will be openly available at \url{https://github.com/mbulla/kilonova\_models} while the data concerning the initial wind configuration for the present simulations are available from the corresponding author, A.P., as well as the hydrodynamic configurations used to extract the light curves, from L.N., upon reasonable request.



\bibliographystyle{mnras}
\bibliography{lnativi} 








\bsp	
\label{lastpage}
\end{document}